\documentclass[aip, reprint, twocolumn, floatfix]{revtex4-2}
\usepackage{float}
\usepackage[utf8]{inputenc}
\usepackage{booktabs}
\usepackage{amsmath}
\usepackage{physics}
\usepackage{graphicx}
\usepackage{gensymb}
\usepackage{amssymb}

\newcommand{\dataTenByDstandoffUm}{1.7}

\newcommand{\dataTenByDstandoffErrorUm}{0.4}

\newcommand{\dataTwentyByDstandoffUm}{1.6}

\newcommand{\dataTwentyByDstandoffErrorUm}{0.1}

\begin{document}

\title{An Integrated Widefield Probe for Practical Diamond Nitrogen-Vacancy Microscopy}

\author{G. J. Abrahams}
\affiliation{School of Physics, University of Melbourne, VIC 3010, Australia}

\author{S. C. Scholten}
\affiliation{School of Physics, University of Melbourne, VIC 3010, Australia}

\author{A. J. Healey}
\affiliation{School of Physics, University of Melbourne, VIC 3010, Australia}
\affiliation{Centre for Quantum Computation and Communication Technology, School of Physics, University of Melbourne, VIC 3010, Australia}

\author{I. O. Robertson}
\affiliation{School of Physics, University of Melbourne, VIC 3010, Australia}

\author{N. Dontschuk}
\affiliation{School of Physics, University of Melbourne, VIC 3010, Australia}

\author{S. Q. Lim}
\affiliation{School of Physics, University of Melbourne, VIC 3010, Australia}

\author{B. C. Johnson}
\affiliation{School of Engineering, RMIT University, Melbourne VIC 3001, Australia}

\author{D. A. Simpson}
\affiliation{School of Physics, University of Melbourne, VIC 3010, Australia}

\author{L. C. L. Hollenberg}
\affiliation{School of Physics, University of Melbourne, VIC 3010, Australia}
\affiliation{Centre for Quantum Computation and Communication Technology, School of Physics, University of Melbourne, VIC 3010, Australia}

\author{J.-P. Tetienne}
\email{jean-philippe.tetienne@rmit.edu.au}
\affiliation{School of Physics, University of Melbourne, VIC 3010, Australia}
\affiliation{Centre for Quantum Computation and Communication Technology, School of Physics, University of Melbourne, VIC 3010, Australia}
\affiliation{School of Science, RMIT University, Melbourne VIC 3001, Australia}


\begin{abstract}
    The widefield diamond nitrogen-vacancy (NV) microscope is a powerful instrument for imaging magnetic fields. However, a key limitation impeding its wider adoption is its complex operation, in part due to the difficulty of precisely interfacing the sensor and sample to achieve optimum spatial resolution.
    Here we demonstrate a solution to this interfacing problem that is both practical and reliably minimizes NV-sample standoff. We built a compact widefield NV microscope which incorporates an integrated widefield diamond probe with full position and angular control, and developed a systematic alignment procedure based on optical interference fringes. Using this platform, we imaged an ultrathin ($1\,{\rm nm}$) magnetic film test sample, and conducted a detailed study of the spatial resolution. We reproducibly achieved an estimated NV-sample standoff (and hence spatial resolution) of at most $\sim 2\,\mu{\rm m}$ across a $\sim 0.5\,{\rm mm}$ field of view. Guided by these results, we suggest future improvements for approaching the optical diffraction limit. This work is a step towards realizing a widefield NV microscope suitable for routine high-throughput mapping of magnetic fields.
\end{abstract}

\maketitle


Magnetic imaging using nitrogen-vacancy (NV) centers in diamond\cite{dohertyNitrogenvacancyColourCentre2013} has proven to be a valuable technique for studying the magnetic and electronic properties of materials and devices\cite{rondinMagnetometryNitrogenvacancyDefects2014,casolaProbingCondensedMatter2018}. One particular approach, widefield NV microscopy, involves imaging an ensemble of NVs residing in a thin ($\lesssim 1\,\mu{\rm m}$) layer within the diamond, whose spin states are read out via their optically detected magnetic resonance (ODMR) spectrum with a camera to map the stray magnetic field of a proximal sample\cite{steinertHighSensitivityMagnetic2010,levinePrinciplesTechniquesQuantum2019,scholten2021widefield}. This technique has found utility in imaging a wide array of devices\cite{nowodzinskiNitrogenVacancyCentersDiamond2015,simpsonMagnetoopticalImagingThin2016,tetienneQuantumImagingCurrent2017,torailleOpticalMagnetometrySingle2018,broadwayImagingDomainReversal2020,turnerMagneticFieldFingerprinting2020,kuImagingViscousFlow2020,meirzadaLongTimeScaleMagnetizationOrdering2021,navaantonioMagneticImagingStatistical2021}, as well as biological\cite{lesageOpticalMagneticImaging2013,glennSinglecellMagneticImaging2015,fescenkoDiamondMagneticMicroscopy2019,mccoeyQuantumMagneticImaging2020} and geological\cite{fuSolarNebulaMagnetic2014,glennMicrometerscaleMagneticImaging2017,fuHighSensitivityMomentMagnetometry2020} samples. While greater spatial resolution is achievable using scanning NV microscopy\cite{rondinMagnetometryNitrogenvacancyDefects2014,casolaProbingCondensedMatter2018}, the parallel operation and high magnetic sensitivity of the widefield modality makes it well-suited to rapid, multi-purpose diagnostic imaging of magnetic materials and devices\cite{scholten2021widefield}. However, its ease of use and potential for high-throughput imaging is often limited in practice by the requirement that the NV layer (and hence diamond sensor) be placed in close proximity with the sample. This is a challenge, as both the sensor and sample are relatively bulky objects (up to millimeters in size). Departures from flatness such as sample surface features or contaminants will cause misalignment between the diamond and sample surfaces resulting in standoffs which can significantly reduce image resolution. Broadly speaking, there are two ways to interface the diamond sensor with the sample\cite{scholten2021widefield}. In the first case, termed ``sample-on-diamond'', the sample is fabricated directly onto the diamond surface. This method is ideal for minimizing standoff (to $10-100$s of nanometers \cite{tetienneQuantumImagingCurrent2017,lesageOpticalMagneticImaging2013,broadwayImagingDomainReversal2020,kuImagingViscousFlow2020}), however it is impractical for measurements of many samples or for samples which cannot be fabricated in-house.  In the second case, termed ``diamond-on-sample'', the diamond sensor is independent from the sample, either placed on the sample or held in contact. This second method partly solves the practicality problem, but makes standoff minimization more difficult, with standoff values as large as $\sim10\,\mu{\rm m}$ frequently reported\cite{torailleOpticalMagnetometrySingle2018,fuHighSensitivityMomentMagnetometry2020,turnerMagneticFieldFingerprinting2020,meirzadaLongTimeScaleMagnetizationOrdering2021}. Here we present a widefield NV microscope with the flexibility of the ``diamond-on-sample'' method, while also demonstrating a systematic method for minimizing standoff.

Our approach follows the method demonstrated by \textcite{ernstPlanarScanningProbe2019}. They reported the application of interferometry to scanning planar probe microscopy, showing that in situations involving a planar sensor parallel to a planar sample, the distance between probe and sample can be reliably minimized by first rotationally aligning the surfaces using interferometry as a feedback mechanism. They implemented this scheme for a $20\,\mu{\rm m}$ sized sample and millimeter-sized sensor, achieving $\approx 20\,{\rm nm}$ standoff at the point of contact, allowing for nanometer-scale imaging through scanning. Here, we extend this concept to larger surface contact areas, as required for widefield imaging of large (millimeter to centimeter) scale samples. 
In the case of widefield imaging, standoff of 100s of nanometers is often sufficient to achieve near-optimum spatial resolution, relaxing the technical requirements. 
Therefore, we implement a simplified and manual version of the alignment method which is suitable in this regime. 

\begin{figure}[t]
    \centering
    \includegraphics[width=1\linewidth]{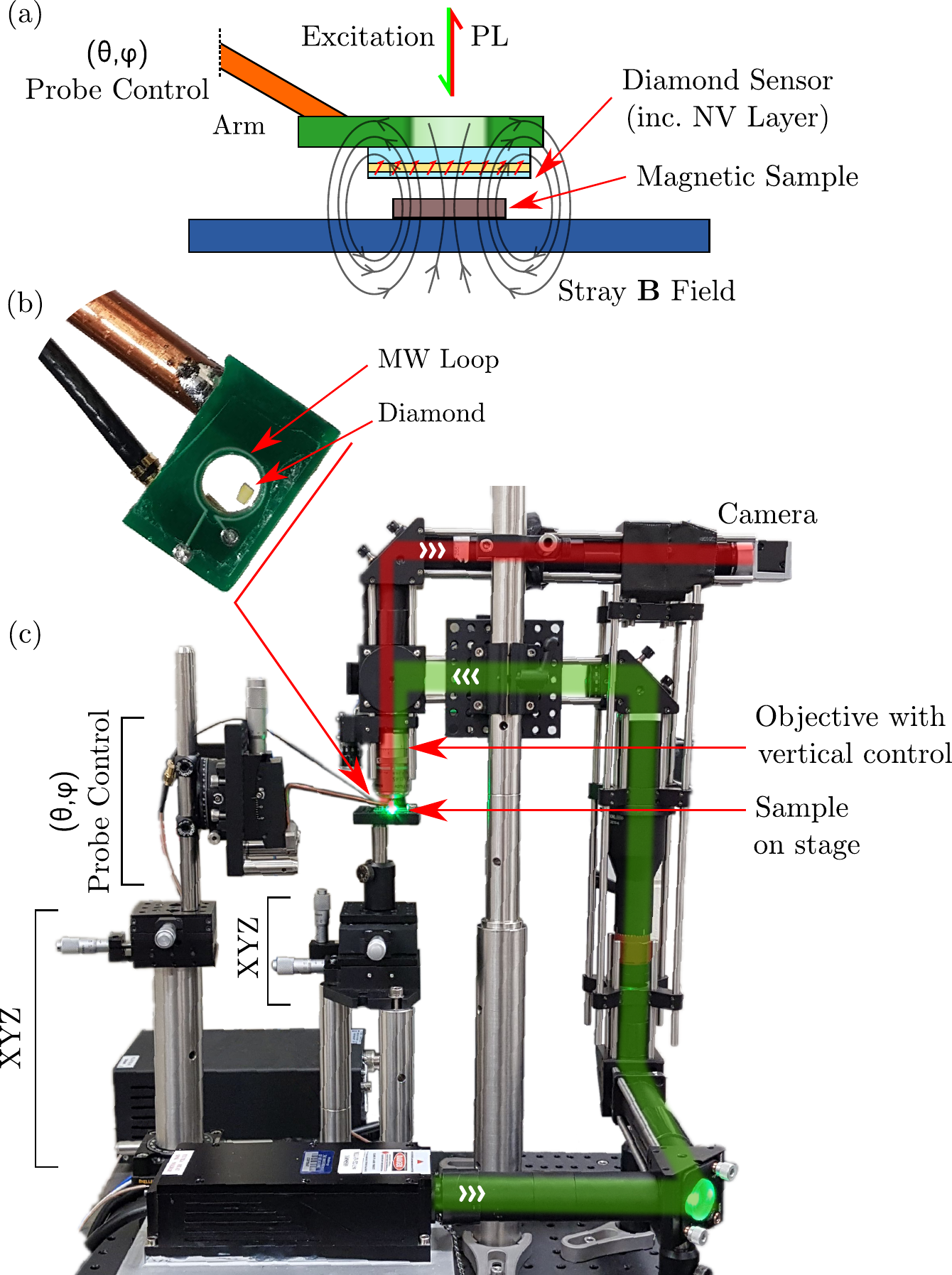}
    \caption{(a) Schematic of the widefield probe concept. A diamond sensor attached to an arm is mechanically brought into alignment with the sample under study. The stray magnetic field is then imaged by the NV layer through its photoluminescence (PL) response. (b) Photograph of the probe, attached to the orientation control arm (copper) and a coaxial cable (black). A microwave (MW) loop antenna is defined onto the PCB. The diamond is glued to a glass coverslip, which is glued to the PCB. (c) Photograph of the complete microscope. Green highlight: $532\,{\rm nm}$ excitation path. Red highlight: $650-750\,{\rm nm}$ PL path. The bias magnet, signal generator and microwave amplifier are not pictured, however all fit onto the breadboard shown.}
    \label{fig:instrument}
\end{figure}

\begin{figure}[t!]
    \centering
    \includegraphics[width=0.9\linewidth]{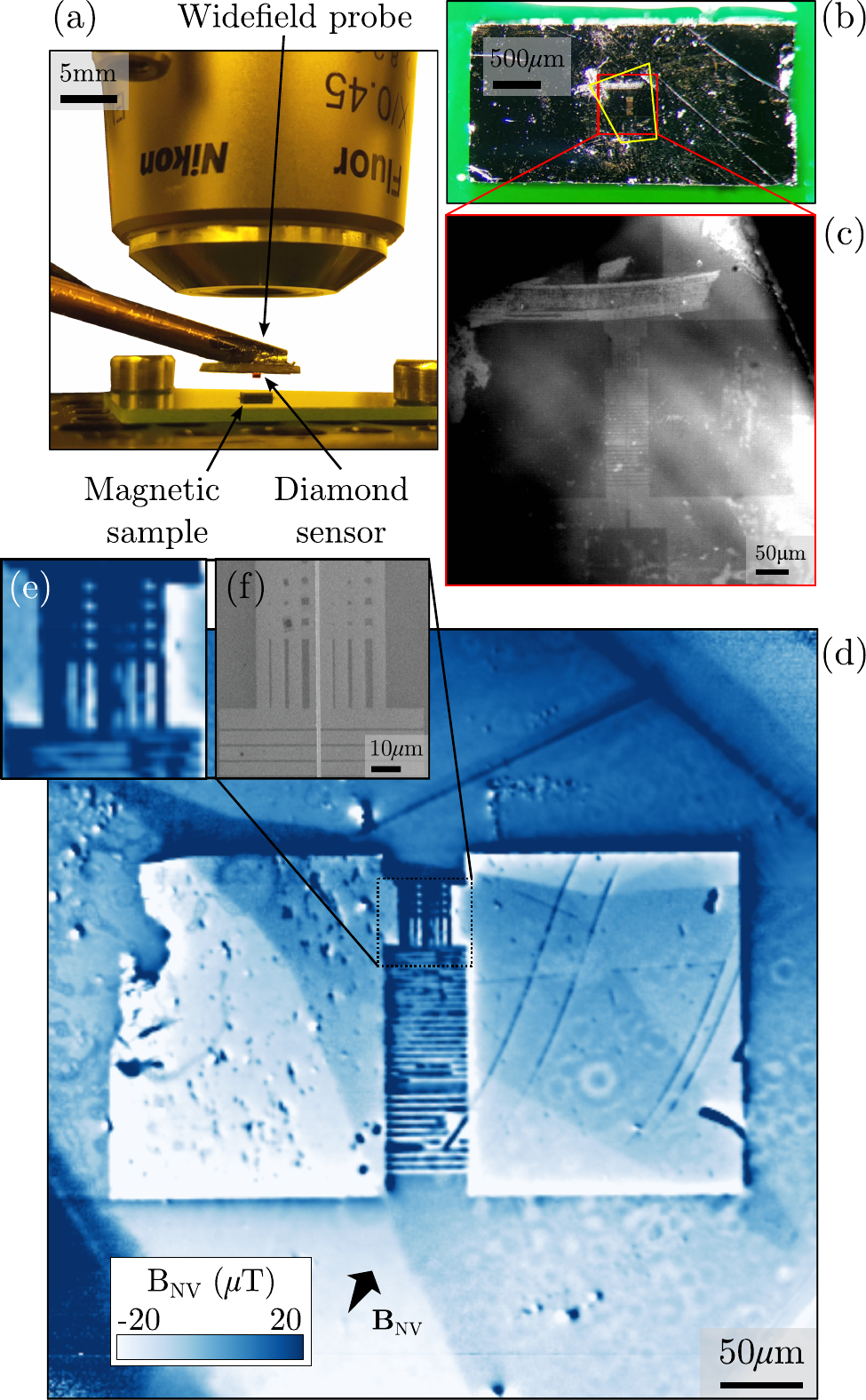}
    \caption{(a) Photograph of the widefield probe before being brought into contact with the sample. (b) Photograph of the sample, sitting atop a bare PCB. Yellow outline indicates approximate diamond footprint. (c) Image of the sample using the reflected NV PL. The diamond is raised to be out of focus; its edges are clearly visible in the top left and top right of the image. The large dark pads on the left and right and the rungs connecting them correspond to the magnetic film, while the rest of the sample is the bare substrate. The smaller square pads on the top and bottom and the line connecting them are the remains of a gold stripline, not used in this experiment. (d) Magnetic map ($B_{\rm NV}$) of the sample after alignment optimization. The diamond position has been moved relative to image (c). The statistical (pixel to pixel) noise is about $1\,\mu{\rm T}$. We applied a 3-point plane subtraction to remove the bias magnetic field. Arrow indicates estimated NV axis direction. (e) Zoom in of the same magnetic image. (f) Scanning Electron Microscope (SEM) image corresponding to (e), showing the smallest device features (dots and vertical bars). From left column to right the features have smallest dimension $0.5$, $1$ and $2$ $\,\mu{\rm m}$ (repeated). }
    \label{fig:device}
\end{figure}

Our platform is based on a ``widefield probe'' which allows angular control of the diamond (the probe) with respect to the fixed sample [Fig. \ref{fig:instrument}(a)]. The widefield probe is constructed on a miniature printed circuit board (PCB) integrating a microwave loop antenna around a through hole (for imaging), and attached to a holding arm [Fig.~\ref{fig:instrument}(b)]. The size of the loop and hole can be varied depending on the application; all measurements here used a $3\,{\rm mm}$ hole diameter. A $\sim0.15\,{\rm mm}$ thick glass coverslip is glued over the through hole, and the diamond is adhered to the coverslip using low fluorescence glue. The arm is attached to a 5-axis stage: 2-axis rotation (rotation + goniometer) configured to allow angular adjustment of the diamond surface with minimal translation, and 3-axis linear motion for positioning and focusing. Finally, this probe stage is integrated into a compact ($45\times45~{\rm cm}^2$) footprint but otherwise standard widefield NV microscope [Fig.~\ref{fig:instrument}(c)]. Long working distance ($\approx 1$\,{\rm cm}) coverslip corrected objective lenses (10x $\rm{NA}=0.3$ and 20x $\rm{NA}=0.45$) are used to give sufficient clearance for the probe, as shown in Fig.~\ref{fig:device}(a).  

To test our method and evaluate integrated system performance, we imaged an ultrathin film, millimeter-sized magnetic device fabricated on a Si/SiO$_2$ substrate [Fig. \ref{fig:device}(b)]. The film is a W/CoFeB/MgO stack with a $1\,{\rm nm}$ thick CoFeB layer exhibiting perpendicular magnetic anisotropy\cite{gross2016}. The diamond used was a $400\,\mu{\rm m}$ thick, $\{100\}$-oriented, high-pressure high-temperature (HPHT) diamond with a $\sim1\,\mu{\rm m}$ NV layer near the surface formed by ion implantation\cite{healeyComparisonDifferentMethods2020} [see further details in Supplementary Section~\ref{sec:diamond}]. A $\sim 1\,{\rm mm}\times0.5\,{\rm mm}$ section of the diamond was laser cut and attached to the widefield probe [Fig.~\ref{fig:instrument}(b)]. 
When focusing the objective onto the sample surface with the diamond at some distance, the device structure is visible in the reflected NV PL [Fig. \ref{fig:device}(c)]. 

\begin{figure}[t]
    \centering
    \includegraphics[width=0.9\linewidth]{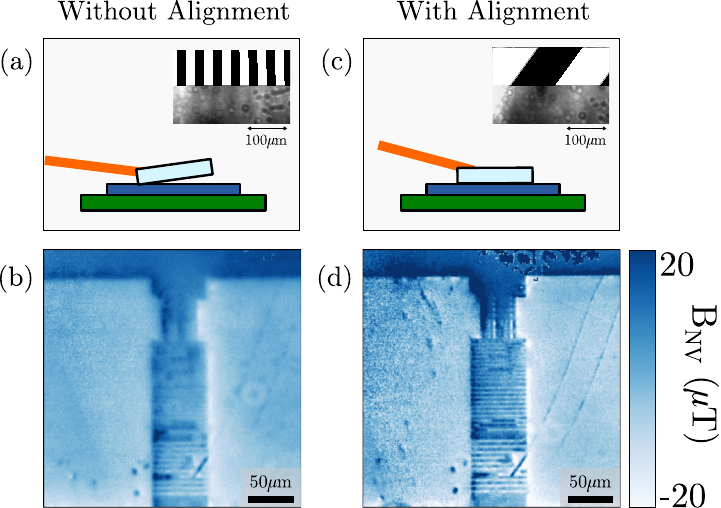}
    \caption{(a,c) Schematic of sensor-sample interfacing without (a) and with (c) angular alignment. Insets depict the fringes seen in the PL image (bottom part is actual data, top part is a guide to the eye). (b,d) Magnetic images corresponding to the situations in (a,c), respectively, showing a clear improvement in sharpness with the alignment step. A plane subtraction to remove the bias magnetic field has been performed.}
    \label{fig:fringes}
\end{figure}

To align the probe with this device, we proceed as follows. As the diamond and sample are coarsely brought together, interference fringes due to angular displacement become faintly visible in the PL image [examples will be shown in Fig.~\ref{fig:fringes}]. The fringes are due to interference of the collimated $532\,{\rm nm}$ excitation beam between diamond and sample surface, resulting in patterned NV excitation and hence modulated PL. Orientation alignment is performed while there is a small air gap between the diamond and sample. Adjusting the diamond angle causes the fringes to rotate and spread along the axis least aligned. Ideally, the fringes are ultimately spread such that only one band is visible.
Finally, the diamond and sample are brought into direct contact by further vertical adjustment. The NV layer is first moved into focus, then the sample is brought up until it is seen to contact the diamond through the NV PL video feed (sample features come into focus, and the sample is stabilized when the diamond comes into contact).
Compared with the method of \textcite{ernstPlanarScanningProbe2019}, this method is manual and does not allow estimation of the actual standoff directly. 

After applying this alignment procedure, we proceed to acquire a magnetic image of the sample. To this end, the NV PL of a region of the diamond is imaged as the microwave frequency is swept, resulting in an ODMR spectrum for each pixel of the image. The sweep is repeated and images are integrated (per frequency bin) to improve the signal to noise ratio. The data is fit numerically to determine the position of the ODMR peaks $f_1$ and $f_2$ of a given NV orientation class, and hence calculate the projected magnetic field $B_{\rm  NV}=(f_2-f_1)/2\gamma_e$, where $\gamma_e=28\,{\rm GHzT}^{-1}$ is the NV gyromagnetic ratio. An example magnetic image is shown in Fig. \ref{fig:device}(d), which reveals device features such as the two large magnetic pads and the rungs connecting them. The relatively clean magnetic image contrasts with the comparatively dirty optical image [Fig. \ref{fig:device}(c)], despite some residual imaging artifacts attributed to diamond features and off-axis bias field alignment [see Supplementary Section~\ref{magimag}]. Note that in Fig. \ref{fig:device}(d) the bias field of $3.6\,{\rm mT}$ was removed via a three-point plane subtraction. The resulting image was obtained using a 10x objective, providing a field of view (FOV) of $565\,\mu{\rm m}$ at $1024\times 1024$ pixels (pixel size $552\,{\rm nm}$). 
Magnetic features such as the rungs were resolved after $\sim 10$ minutes of integration. To resolve micron-scale details required integration times on the order of hours. 
As seen in the zoom-in magnetic image and scanning electron microscope image [Fig.~\ref{fig:device}(e-f)], magnetic bars down to a width of $0.5\,\mu{\rm m}$ can be resolved.
All resolved features are separated by $\approx 5\,\mu{\rm m}$ implying that the resolution is at or below $5\,\mu{\rm m}$.

To illustrate the importance of the alignment step, we show images resulting from two different imaging conditions, obtained using a 20x objective (hence, resulting in a smaller FOV compared to Fig.~\ref{fig:device}(d)). In the first case, no angular alignment is performed [Fig. \ref{fig:fringes}(a)], and fringes are visible in PL [Inset to (a) and Supplementary Fig.~\ref{fig:fringeshighlighted}(a)]. As a result, the magnetic image is blurry [Fig. \ref{fig:fringes}(b)]. We found that forcing the diamond into a closer contact without first performing alignment does not actually reduce the standoff,  which can be understood by considering the schematic in Fig. \ref{fig:fringes}(a), where the diamond contacts the sample at a point resulting in non-uniform standoff. 
We also ``dropped'' the diamond onto the sample and took ODMR images with microwaves provided by a PCB. This produced similarly blurry magnetic images. This is likely due to either sample surface features or contaminants causing similar point like contact, resulting in the diamond resting at an angle.
In contrast, minimizing the angle through monitoring the fringes before lowering the diamond reliably produced the sharpest images [Fig. \ref{fig:fringes}(c,d)]. This is because the aligned diamond will overall be within close proximity with the sample. Alignment is maintained while contact is made, thus the aforementioned contact points may be forced away (i.e. contaminants) or otherwise only limit overall standoff to a small distance, rather than causing large standoff.
The relative tilt between sensor and sample plane is determined by the fringe separation. The full PL FOV for the 10x objective is $1130\,\mu{\rm m}\times 1130\,\mu{\rm m}$ (this is cropped by a factor $2$ for the magnetic image in Fig.~\ref{fig:device}(d)). Thus, once a single fringe is visible, the relative tilt between sensor and sample planes is within $0.47\,{\rm mrad}$. This corresponds to $\sim 0.5\,\mu{\rm m}$ variation in standoff across the $\sim 1\,{\rm mm}$ FOV. We found this precision to be sufficient for noticeable improvement compared to not performing alignment.

\begin{figure}[t]
    \centering
    \includegraphics[width=0.95\linewidth]{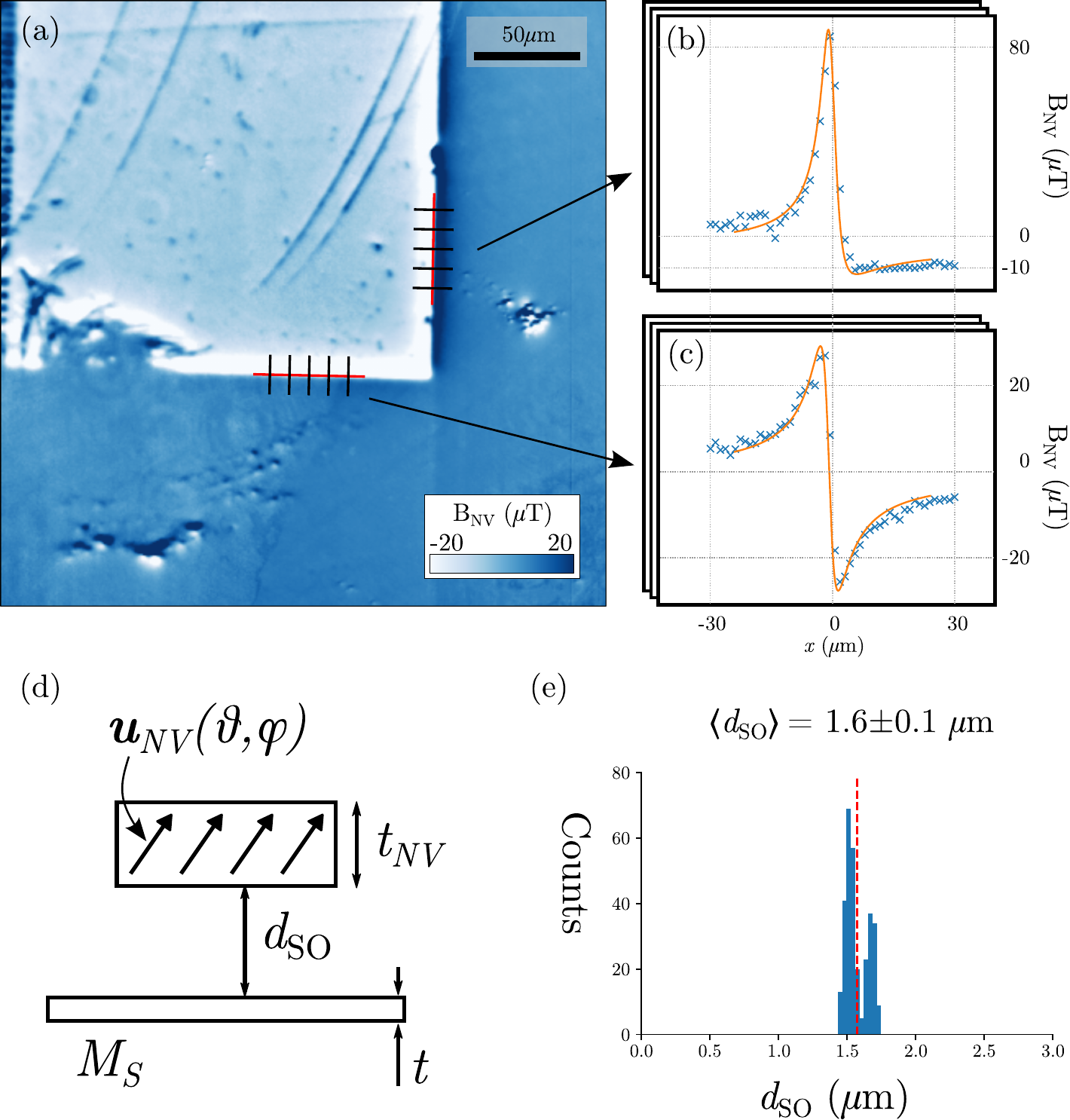}
    \caption{(a) Magnetic image centered on a corner of the pad. Note some additional features compared to Fig.~\ref{fig:device}(d), indicating damage to the magnetic film has occurred as a result of repeated cleaning. Line-cuts (black) are taken along segments of edges (red). (b,c) Example line-cuts along perpendicular directions. Orange lines are the magnetic model fit. (d) Diagram indicating model parameters. (e) Histogram of $d_{\rm SO}$ parameter fits to line-cut data. The red line indicates the mean value.}
    \label{fig:magneticres}
\end{figure}

Having demonstrated a reliable way to align the NV-sample interface, we now investigate the factors affecting spatial resolution in the magnetic images. Firstly there are `optical' effects such as diffraction, and also aberrations due to imaging through the diamond and coverslip. These effects limit the resolution in imaging the PL of the NV layer, independent of sample. Secondly, there are `standoff' effects, namely the fact that the stray magnetic field seen by the NV layer is a convolution of the sample's magnetization with a resolution function whose width is proportional to the NV-sample standoff $d_{\rm SO}$\cite{limaObtainingVectorMagnetic2009,casolaProbingCondensedMatter2018}, as well as averaging over the finite thickness of the NV layer\cite{broadwayImprovedCurrentDensity2020}.
To estimate $d_{\rm SO}$, we follow the method in \textcite{hingant2015}, analyzing line-cuts over the edge of the large magnetic pads [Fig.~\ref{fig:magneticres}(a-c)]. The model is given by the theoretical stray field at a free standoff $d_{\rm SO}$, integrated over the NV layer thickness, and convolved with a Gaussian function of fixed width $r_{\rm opt}$ to capture optical effects. The other free parameters are the NV orientation [spherical coordinates $(\theta,\phi)$] and sample spontaneous magnetization $M_s$. The assumed parameters are the film thickness ($t=1\,{\rm nm}$) and NV layer thickness, approximated to be $t_{\rm NV}=1\,\mu{\rm m}$ with a uniform distribution [Fig.~\ref{fig:magneticres}(d)].
Although in principle the optical resolution may be worse (larger profile width) than the diffraction limit, we approximate them to be equal, which resulted in a good model fit. This leads to the estimated standoff $d_{\rm SO}$ being an upper bound for the actual standoff. 
The analysis of Fig.~\ref{fig:magneticres}(a), which was obtained using the 20x objective, leads to $d_{\rm SO}=\dataTwentyByDstandoffUm\pm\dataTwentyByDstandoffErrorUm\,\mu{\rm m}$ [Fig.~\ref{fig:magneticres}(e)].
Similarly, from the image in Fig.~\ref{fig:device}(d) which used the 10x objective, we find $d_{\rm SO}=\dataTenByDstandoffUm\pm\dataTenByDstandoffErrorUm\,\mu{\rm m}$. Such physical standoffs are similar to the best values reported by dropping the diamond and manual trial and error\cite{bertelliMagneticResonanceImaging2020,simpsonMagnetoopticalImagingThin2016}, and significantly improved compared to more typical values on the order of $10\,\mu{\rm m}$\cite{torailleOpticalMagnetometrySingle2018,fuHighSensitivityMomentMagnetometry2020,turnerMagneticFieldFingerprinting2020,meirzadaLongTimeScaleMagnetizationOrdering2021}.
Together, the standoff and optical resolution combine to give an effective spatial resolution for a magnetic edge of $\approx 4\,\mu{\rm m}$ [See Supplementary Section~\ref{standoff}]. This value is dominated by the effect of standoff, as the stray field over an edge has a width of $2\times d_{\rm SO}$\cite{tetienneNatureDomainWalls2015}. Including NV layer thickness, the contribution due to standoff is roughly $4\times$ that of diffraction, hence improvements to the effective resolution should first involve further minimization of $d_{\rm SO}$, followed by improvements to $r_{\rm opt}$ and finally the diffraction limit. Note that for a magnetic dipole, the stray field width is approximately equal to $d_{\rm SO}$ [See Supplementary Section~\ref{magneticres}], giving a resolution of $\approx 2\,\mu{\rm m}$ in our case when trying to resolve dipole-like magnetic objects. 
From the model fit, we also obtain $\theta$ and $M_s$, finding they are broadly consistent with expectations [See Supplementary Section \ref{magneticres}].

Overcoming standoff limitations will require finer angular and vertical control of the probe, easily accomplished using robotic stages\cite{ernstPlanarScanningProbe2019}. Keeping the surfaces clean and flat to further reduce $d_{\rm SO}$ should be readily achievable.
Neglecting NV layer thickness, to attain an effective resolution roughly $1.2\times$ the diffraction limit for a $\rm{NA}=0.45$ objective requires the standoff $d_{\rm SO}$ to be $\sim 150\,{\rm nm}$.
To ensure the NV layer thickness is not limiting, diamonds with a thinner ($\lesssim100$\,nm) layer should be used, at the expense of a reduced magnetic sensitivity\cite{healeyComparisonDifferentMethods2020}. Secondly, diffraction-limited optical resolution should be attainable by using a thinner diamond or customized optical components to minimize aberrations. A final resolution improvement can be achieved by lowering the diffraction limit, down to $350$\,nm using an objective with $\rm{NA}=0.9$. This will require some optimization of the widefield probe design to accommodate the millimeter-scale working distance of such high-NA objectives.

In summary, we demonstrated an NV microscope based on an integrated widefield NV probe which provides a rapid and reliable way to interface the probe with the sample, making our system ideal for high-throughput measurements of a range of magnetic samples and electronic devices. From our analysis of spatial resolution, we found the effective spatial resolution to be dominated by standoff. With further optimization of the spatial resolution and automation of the alignment procedure, there is an excellent prospect for this instrument to become a routine magnetic imaging technique.

\begin{acknowledgments}
We thank V. Jacques for providing the CoFeB sample, which was originally grown and patterned by M. Hayashi, K. Garcia and J.-P. Adam. This work was supported by the Australian Research Council (ARC) through grants CE170100012, DP190101506 and FT200100073. We acknowledge the AFAiiR node of the NCRIS Heavy Ion Capability for access to ion-implantation facilities. S.C.S gratefully acknowledges the support of an Ernst and Grace Matthaei scholarship. A.J.H. is supported by an Australian Government Research Training Program Scholarship.

\end{acknowledgments}

\section*{Data Availability Statement}

The data that support the findings of this study are available from the corresponding author upon reasonable request.

\section*{Author Declarations}

The authors have no conflicts to disclose.

\section*{References}

\bibliography{refs}

\clearpage

\renewcommand{\theequation}{S\arabic{equation}}
\renewcommand{\thefigure}{S\arabic{figure}}
\renewcommand{\thetable}{S\arabic{table}}
\setcounter{figure}{0}

\begin{widetext}

\section*{Supplementary Information}

\section{NV-diamond sensor}
\label{sec:diamond}

The NV-diamond samples used in this work were made from a $4\,{\rm mm}\times 4\,{\rm mm}\times 400\,\mu{\rm m}$ type-Ib, single-crystal diamond substrate grown by high-pressure, high-temperature (HPHT) synthesis, with $\{100\}$-oriented polished faces, purchased from Chenguang Machinery \& Electric Equipment. The diamond featured multiple sectors with various levels of nitrogen content as inferred from the different shades of yellow [Fig.~\ref{Fig_SRIM}(a)], but all sectors had very low level of NV PL initially. 

To form a dense NV layer near the surface, we implanted the diamond with $4$\,MeV Sb ions with a dose of $3\times 10^{11}$ ions/cm$^2$. We performed full cascade Stopping and Range of Ions in Matter (SRIM) Monte Carlo simulations to estimate the depth distribution of the created vacancies [Fig.~\ref{Fig_SRIM}(b)], predicting a distribution spanning the range $0$~-~$1100$\,nm (approximated to a $1~\mu$m layer in the main text) with a peak vacancy density of $\sim170$~ppm at a depth of $\sim 800$\,nm (assuming no dynamic annealing effects). Following implantation, the diamond was annealed at $900$\,$^\circ$C for 4 hours in a vacuum of $\sim10^{-5}$~Torr to form the NV centers. After annealing, the plate was acid cleaned ($30$\,minutes in a boiling mixture of sulphuric acid and sodium nitrate).

We then took a photoluminescence (PL) map [Fig.~\ref{Fig_SRIM}(c)] to identify the brightest sectors, as these gave the best magnetic sensitivity. These brightest sectors generally corresponded to the sectors with the lightest shade of yellow, indicating that the darker sectors have a high level of nitrogen such that electron tunnelling significantly quenches the PL\cite{Capelli2019,mansonNVPairCentre2018}. We then used a laser cutting system to isolate the best sectors [Fig.~\ref{Fig_SRIM}(d)], resulting in diamonds of lateral size $<1\,{\rm mm}^2$ each [Fig.~\ref{Fig_SRIM}(e)]. The diamond pieces were acid cleaned (same process as above) after laser cutting, leaving them ready for mounting on the widefield probe as described in the main text. A photo of such a probe as seen from above (objective side) is shown in Fig.~\ref{Fig_SRIM}(f). 

\begin{figure}[ht]
	\begin{center}
		\includegraphics[width=1\columnwidth]{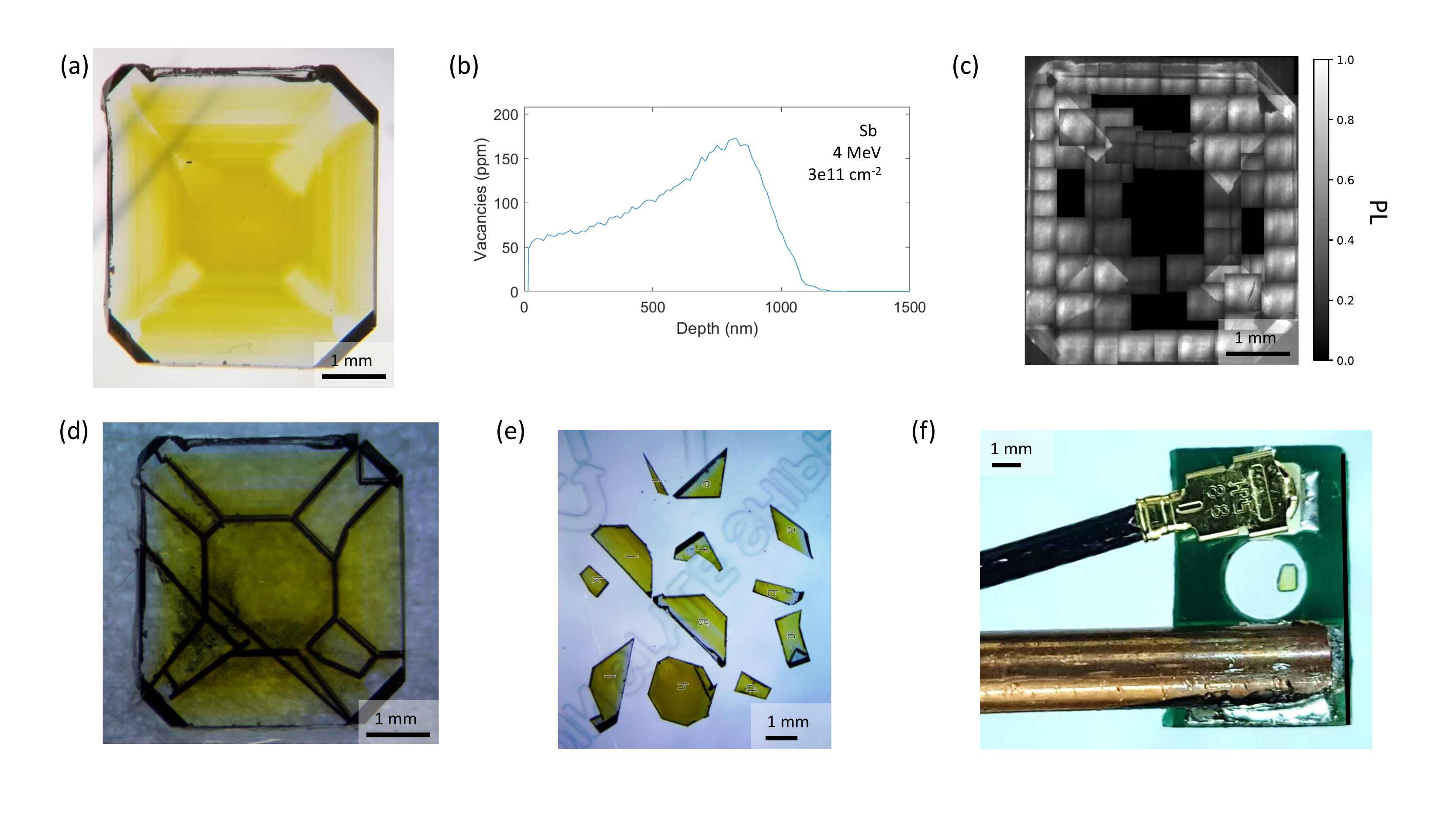}
		\caption{(a) Photograph of the diamond as purchased. (b) Stopping and Range of Ions in Matter (SRIM) simulation of the implantation used, showing the vacancy concentration as a function of depth. We assumed a diamond density of $3.51$\,g\,cm$^{-3}$ and a displacement energy of $50$\,eV. (c) Stitched PL map of the diamond shown in (a) after implantation and annealing. (d) Photograph of the same diamond after laser cutting. (e) Photograph of the resulting pieces after separation and acid cleaning. (f) Photograph of the probe from above. Diamond is visible in the right half of the thru-hole.}
		\label{Fig_SRIM}
	\end{center}
\end{figure}

\section{Experimental setup}

The experimental setup is shown in Fig. 1(c) of the main text. Continuous-wave (CW) laser excitation was produced by a $\lambda = 532$~nm solid-state laser outputting 1.5 W of power (Dragon Lasers FN series). The beam was attenuated to 300 mW, beam expanded (10x) and focused using a widefield lens ($f = 100\,{\rm mm}$) to the back aperture of the objective lens. We used two different objectives in this work: 
(i) Nikon S Plan Fluor 20x ELWD, NA = 0.45, working distance 6.9-8.2~mm; (ii) Nikon Plan Fluor 10x, NA = 0.3, working distance 16~mm. The PL from the NV layer was collected by the same objective, separated from the excitation light with a dichroic mirror, filtered using a longpass filter, and imaged using a tube lens ($f = 200$ mm) onto a CMOS camera (Basler acA2040-90um USB3 Mono). 

Microwave excitation was provided by a signal generator (Windfreak SynthNV Pro) gated using a switch (Mini-Circuits ZASWA-2-50DR+) and amplified (Mini-Circuits ZHL-16W-43+). The output of the amplifier is directly connected to the PCB of the widefield probe using a coaxial cable. A pulse pattern generator (SpinCore PulseBlasterESR-PRO 400 MHz) was used to gate the microwave and to synchronise the image acquisition. 

The optically detected magnetic resonance (ODMR) spectra of the NV layer were obtained by sweeping the microwave frequency, taking a reference image (microwave off) for each frequency in order to produce a normalized spectrum removing common-mode PL fluctuations. The exposure time for each camera frame was 20~ms (i.e. 40~ms per frequency), and the entire sweep was repeated thousands of times to improve the signal to noise ratio. A bias magnetic field of a few milliteslas was applied using a permanent magnet, roughly aligned (but not exactly) with one of the four possible NV orientation classes (corresponding to the $\langle 111\rangle$ crystal directions). We swept the microwave frequency in order to interrogate this mostly aligned NV family only. All measurements were performed in an ambient environment at room temperature.

\section{Alignment using PL}

As mentioned in the main text, alignment is performed using optical fringes. The full PL images corresponding to the two different alignment conditions in main text Fig.~\ref{fig:fringes} are shown in Fig.~\ref{fig:fringeshighlighted}. The insets in main text Figs.~\ref{fig:fringes}(a,c) are sub-regions of these full PL images (with a $90^\circ$ rotation) selected to better highlight the fringes, with black/white stripes to guide the eye. While the interference fringes are quite difficult to see in these static images, they are relatively easy to identify in the live video feed as the angle is adjusted, enabling optimization.  

The fringe spacing corresponds to the relative angle between the sensor and sample surfaces: $\theta\approx\lambda/\Delta$ where $\theta$ is the angular displacement, $\lambda$ is the excitation wavelength ($532\,{\rm nm}$) and $\Delta$ is the fringe spacing.

\begin{figure}[h!]
    \centering
    \includegraphics[width=0.5\textwidth]{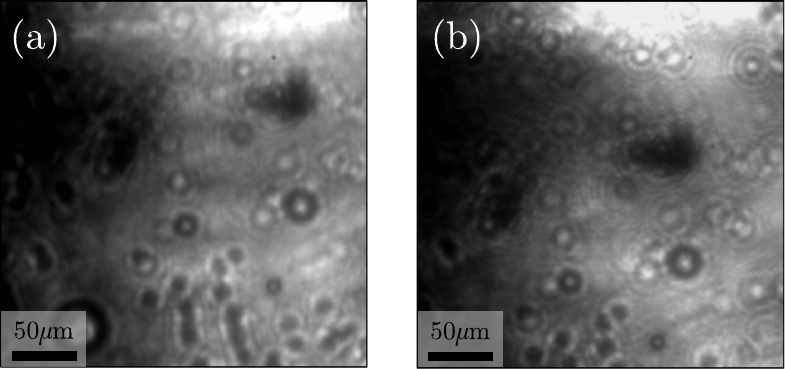}
    \caption{(a,b) Full NV PL images corresponding to the magnetic images in Figs.~\ref{fig:fringes}(b,d) of the main text, respectively. Interference fringes are (faintly) visible in (a), and they become more separated in (b) indicating a smaller angle between diamond and sample. In both images the fringes are quite difficult to see, however in the live video feed they are more apparent as the angle is varied, enabling optimization.}
    \label{fig:fringeshighlighted}
\end{figure}

\section{Magnetic field imaging\label{magimag}}

The ODMR peaks $f_{1,2}$ are generally dependent on  temperature, strain and magnetic field\cite{dohertyNitrogenvacancyColourCentre2013}. In the regime where the applied field along the NV axis is large compared to that perpendicular, $f_{1,2}$ are mostly dependent on magnetic field\cite{rondinMagnetometryNitrogenvacancyDefects2014},
\begin{equation}
f_{1,2} = D \mp \gamma_e B_{\rm NV}
\end{equation} 
where $D=2.87\,{\rm GHz}$ is the zero field splitting, $\gamma_e=28\,{\rm GHzT}^{-1}$ is the gyromagnetic ratio and $B_{\rm NV}$ is the magnetic field projection along the NV axis.

In this aligned field regime, the magnetic field can then be deduced from the two ODMR frequencies,
\begin{equation} \label{eq:BNV}
B_{\rm NV} = \frac{f_2-f_1}{2\gamma_e}. 
\end{equation} 
The $B_{\rm NV}$ maps in the main text were obtained in this way. Additionally, the zero-field splitting $D$ can be obtained via 
\begin{equation} \label{eq:D}
D = \frac{f_1+f_2}{2}. 
\end{equation} 
The $D$ map corresponding to the data in main text Fig.~\ref{fig:device}(d) is shown in Fig.~\ref{fig:dmap}. Large uniform sectors with different $D$ values are visible, which match sectors with different levels of nitrogen content in our HPHT diamond. The variation in $D$ between different sectors is attributed to differences in strain.

Comparing Fig.~\ref{fig:dmap}(b) (which is the same as Fig.~\ref{fig:device}(d) of the main text but without 3-point plane subtraction, and nominally representing $B_{\rm NV}$) with Fig.~\ref{fig:dmap}(a) (nominally representing $D$), we notice some correlations: the strain sectors are visible in the $B_{\rm NV}$ map, whereas the magnetic devices are visible in the $D$ map. This cross-talk is due to the fact that here the bias magnetic field was not well aligned with the NV axis. In the misaligned regime, Eq. \ref{eq:BNV} and \ref{eq:D} are only approximate relations\cite{rondinMagnetometryNitrogenvacancyDefects2014}, leading to an apparent cross-talk between the resulting maps. However, more accurate alignment or performing 8-peak ODMR (measuring all 4 NV orientation classes\cite{broadwaySpatialMappingBand2018}) can reliably resolve this. For example, the magnetic images shown in Figs. \ref{fig:fringes} and \ref{fig:magneticres} of the main text were obtained with a better bias field alignment, leading to significantly reduced strain-induced artifacts.

\begin{figure}[h!]
    \centering
    \includegraphics[width=0.9\textwidth]{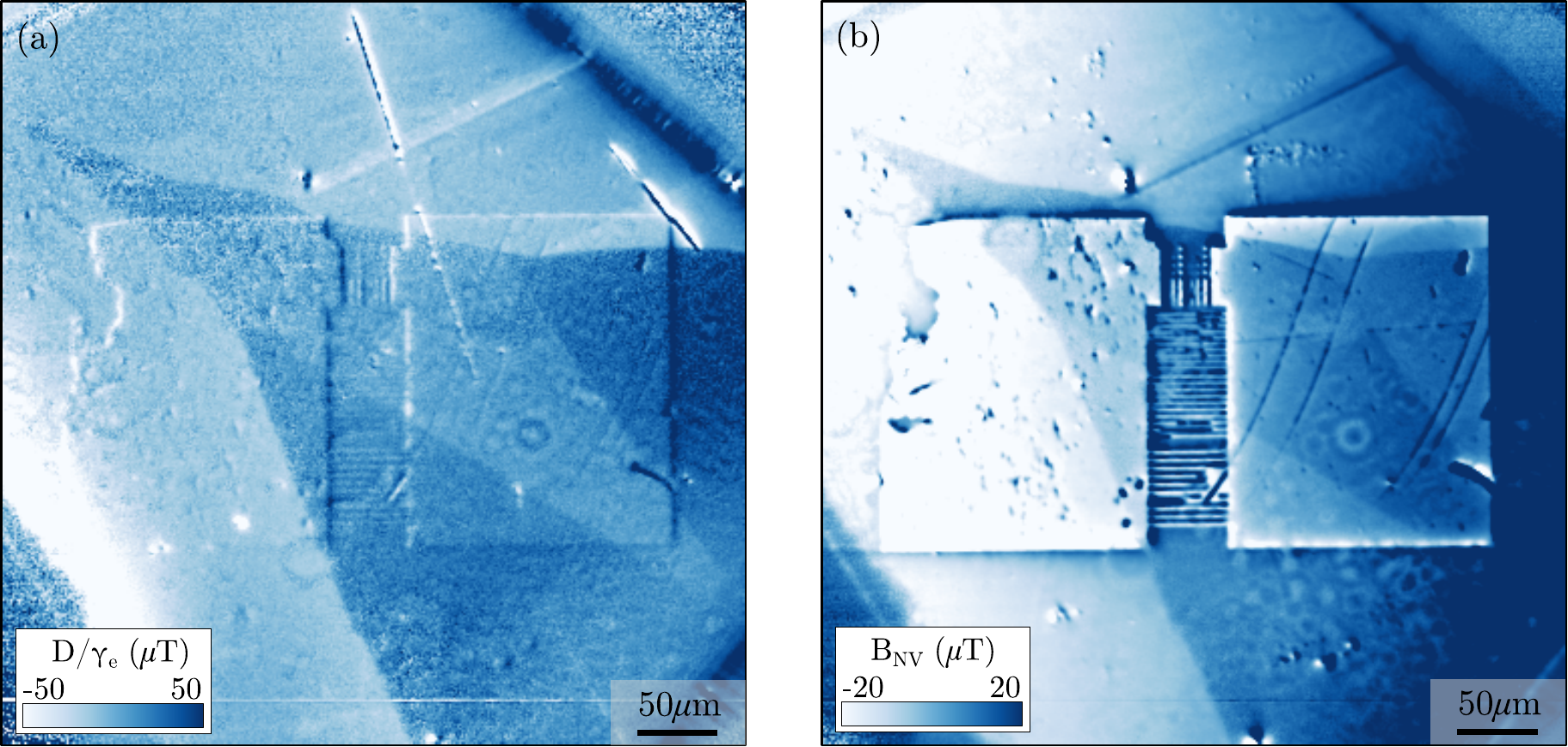}
    \caption{Images obtained from the same data that led to Fig.~\ref{fig:device}(d) of the main text. (a) Map of the zero-field splitting parameter $D$ deduced from Eq.~\ref{eq:D}. (b) Raw map of $B_{\rm NV}$ as deduced from Eq.~\ref{eq:BNV}, without 3-point plane subtraction.
    }
    \label{fig:dmap}
\end{figure}

\section{Magnetic field fitting\label{magneticres}}

The magnetic field along a line-cut perpendicular to an infinite edge of a perpendicularly magnetized film (spontaneous magnetization $M_s$, thickness $t$) at standoff distance $d_{\rm SO}$ is given by\cite{hingant2015}:
\begin{align}
    B_x &= \frac{\mu_0 M_s}{4\pi}\ln\left(\frac{x^2+(d_{\rm SO}+\frac{t}{2})^2}{x^2+(d_{\rm SO}-\frac{t}{2})^2}\right)\label{eq:Bx}\\
    B_z &= \frac{\mu_0 M_s}{2\pi}\left[\arctan\left(\frac{x}{d_{\rm SO}+\frac{t}{2}}\right)-\arctan\left(\frac{x}{d_{\rm SO}-\frac{t}{2}}\right)\right]\label{eq:Bz}
\end{align}
where $x$ is along the line-cut (film edge at $x=0$) and $z$ is normal to the magnet plane. The magnetic field projection along the NV axis is then
\begin{align}
    B_{\rm NV} = \sin\theta\cos\phi B_x + \cos\theta B_z
    \label{eq:BNV2}
\end{align}
where $(\theta,\phi)$ are the spherical angles characterizing the NV axis direction.

Taking Eq. \ref{eq:Bx} to the thin film limit ($t\ll  d_{\rm SO}$) gives a Lorentzian profile, with a full width at half maximum (FWHM) equal to $2\times d_{\rm SO}$\cite{hingant2015}. This profile is the result of upwards continuation of the magnetic field from the sample plane (a delta function in this case for the $B_x$ component) to the sensing (NV) plane\cite{limaObtainingVectorMagnetic2009}. Thus, one can view the Lorentzian of width $2d_{\rm SO}$ as the point spread function (PSF) associated with the stray field measurement at a finite standoff, which is a helpful concept when discussing spatial resolution. In general, however, the PSF may differ depending on the type of magnetic field source and the field component considered\cite{casolaProbingCondensedMatter2018}. For example, in the case of a magnetic dipole oriented in the $z$-direction, the profile of the $B_z$ component is narrower, having a FWHM approximately equal to $d_{\rm SO}$. Nevertheless, the case of a film edge is of relevance for many applications for instance to 2D magnetic materials\cite{thielProbingMagnetism2D2019,broadwayImagingDomainReversal2020}. Therefore, in the following discussion we take the effective spatial resolution (considering standoff effects only) to be $r_{\rm eff}=2d_{\rm SO}$.

To account for optical effects, Eq.~\ref{eq:BNV2} in principle needs to be convolved with the optical PSF, assumed to be a Gaussian profile with FWHM $d_{\rm opt}$. As mentioned in the main text, we simply took $d_{\rm opt}=d_{\rm diff}$ where $d_{\rm diff}$ corresponds to the diffraction limit, which potentially results in an overestimate of $d_{\rm SO}$. In future we plan to characterize $d_{\rm opt}$ more precisely.
To account for a finite NV layer thickness $t_{\rm NV}$, $d_{\rm SO}$ in Eq.~\ref{eq:BNV2} is replaced with $z$, and the average is taken over $d_{\rm SO}\leq z \leq d_{\rm SO}+t_{\rm NV}$.
Thus:
\begin{align}
    B_{\rm NV}^* &= \frac{1}{t_{\rm NV}}\int_{d_{\rm SO}}^{d_{\rm SO}+t_{\rm NV}}dz B_{\rm NV}(x, z) \circledast {\rm Gauss}(d_{\rm opt})\label{BNVstar}
\end{align}

We assume the magnet to have uniform magnetization and thickness ($t=1\,{\rm nm}$). 
We also assume the standoff to be uniform across the image after alignment, hence $d_{\rm SO}$ is uniform. Actually, at alignment, no fringes are visible, hence fringe separation may be taken to be the full image width. This implies the maximum standoff variation between one edge and another is $1$ wavelength, i.e. $0.7\,\mu{\rm m}$. Thus the assumption that standoff is uniform amounts to estimating the average standoff across the image, whereas it would also be possible to map the variable standoff along the edges of the device.
Simultaneously fitting pairs of orthogonal line-cuts extracted from two perpendicular edges (edge 1 and edge 2) [Fig. \ref{fig:magneticres}(a)], we can recover the magnetization $M_s$, the NV angles $\theta$ and $\phi$, and the standoff $d_{\rm SO}$. Fitting is performed using non-linear least squares (using SciPy\cite{virtanen2020}), with the residual function $\epsilon$ of each pair given by:
\begin{align}
    \epsilon &= \sum_{i}\left[B_{\rm NV}^*(x_{i,1}-x^0_1, \theta, \phi) + c_1 - y_{i,1}\right]^2 + \sum_{i}\left[B_{\rm NV}^*\left(x_{i,2}-x^0_2, \theta, \phi+\frac{\pi}{2}\right) + c_2 - y_{i,2}\right]^2
\end{align}
where $x_{i,j}$ and $y_{i,j}$ are the $i^{th}$ coordinate and data point of the $j^{th}$ line-cut respectively ($j\in\{1,2\}$). The parameters $x^0_j$ and $c_j$ are to center the fits to the data horizontally and vertically respectively.
A histogram is constructed for each fit parameter, from which we calculate the measured value (mean) and an error estimate ($\pm$ one standard deviation), which are the numbers discussed in the main text and given in  Table~\ref{tab:results}.
%
%
We measured $\theta=55\pm5\degree$ (10x objective) and $\theta=53\pm2\degree$ (20x), consistent with the expectation for a $\{100\}$-oriented diamond (nominally $\theta=54.7\degree$).
We also measured $M_s=0.6\pm0.1\,{\rm MAm}^{-1}$ (10x) and $M_s=1.03\pm0.04\,{\rm MAm}^{-1}$ (20x), consistent with the previous measurement of the same sample with scanning NV magnetometry\cite{gross2016}, $M_s=0.82\pm0.10\,{\rm MAm}^{-1}$. 
Not addressed in this work is the effect of delocalized contributions to the ODMR spectrum of each pixel due to reflections of PL internal to the diamond. In their experiment, Fu et al. estimated this background signal to contribute $50\%$ of measured spectra\cite{fuHighSensitivityMomentMagnetometry2020}. Applying the same factor here would result in a doubling of the value of $M_s$ measured. However, systematically determining the ratio of delocalized to local signal was beyond the scope of this experiment, and will be addressed in future work.

\begin{table}[]
    \centering
    \begin{tabular}{ccccc}
        \toprule
        Objective & $d^*_{\rm SO}$ & $d_{\rm SO}$ & $M_s$ & $\theta$ \\
         & ($\,\mu{\rm m}$) & ($\,\mu{\rm m}$) & ($\,{\rm MAm}^{-1}$) & ($\degree$)\\
        \midrule
         10x & $2.20\pm0.40$ & $1.73\pm0.41$ & $0.62\pm0.09$ & $55.30\pm5.33$\\
         20x & $2.03\pm0.09$ & $1.57\pm0.08$ & $1.03\pm0.04$ & $52.61\pm1.70$\\
        \bottomrule
    \end{tabular}
    \caption{Summary of the values obtained for each objective used. $d_{\rm SO}^*$} is the standoff found when the NV layer thickness is neglected.
    \label{tab:results}
\end{table}

\section{Optical and Magnetic Resolution\label{standoff}}

As discussed, resolution can be characterized by three operations: convolution with a Gaussian due to optical effects, convolution with a Lorentzian due to standoff, and averaging over the NV layer thickness. Note that the Gaussian is an approximation to the diffraction PSF which is an Airy disk, but we used a Gaussian as a more generic optical PSF, allowing us to test the scenario of a PSF broadened by aberrations, $d_{\rm opt}>d_{\rm diff}$. However, we found the $d_{\rm opt}=d_{\rm diff}$ assumption to give the best fit to the data. To determine the effective resolution mentioned in the main text, we find the FWHM of an apparent PSF. This apparent PSF is given by the convolution of the Gaussian and Lorentzian PSFs, neglecting the averaging over the NV layer for simplicity. The resulting function is the Voigt profile. To account for the NV layer thickness, the Lorentzian PSF FWHM $2d_{\rm SO}$ is replaced with FWHM $2d_{\rm SO}^*$, where $d_{\rm SO}^*$ is the apparent standoff determined by fitting the line-cuts to the data without including the averaging of the NV layer [See Table~\ref{tab:results}]. As expected, we find $d_{\rm SO}^*\approx d_{\rm SO}+t_{\rm NV}/2$. The FWHM of a Voigt profile $r_V$ composed of a Gaussian of FWHM $r_G$ and Lorentzian of FWHM $r_L$ is well approximated by\cite{olivero1977}: 
\begin{equation}
    r_V = 0.5346r_L + \sqrt{0.2166r_L^2+r_G^2}
\end{equation}

Thus the effective resolution $r_{\rm eff}$ is given by:
\begin{equation}
    r_{\rm eff} = 0.5346\times 2d_{\rm SO}^* + \sqrt{0.2166\times(2d_{\rm SO}^*)^2+r_{\rm opt}^2}
\end{equation}

To illustrate the next steps in improving resolution, we graphed the NV-sample standoff against effective resolution for objectives with different numerical apertures [Fig.~\ref{fig:effectivestandoff}]. The measurements made in this work using the 10x and 20x objectives are shown. At zero standoff, the effective resolution is equal to the diffraction limited resolution. 

\begin{figure}
    \centering
    \includegraphics[width=0.7\textwidth]{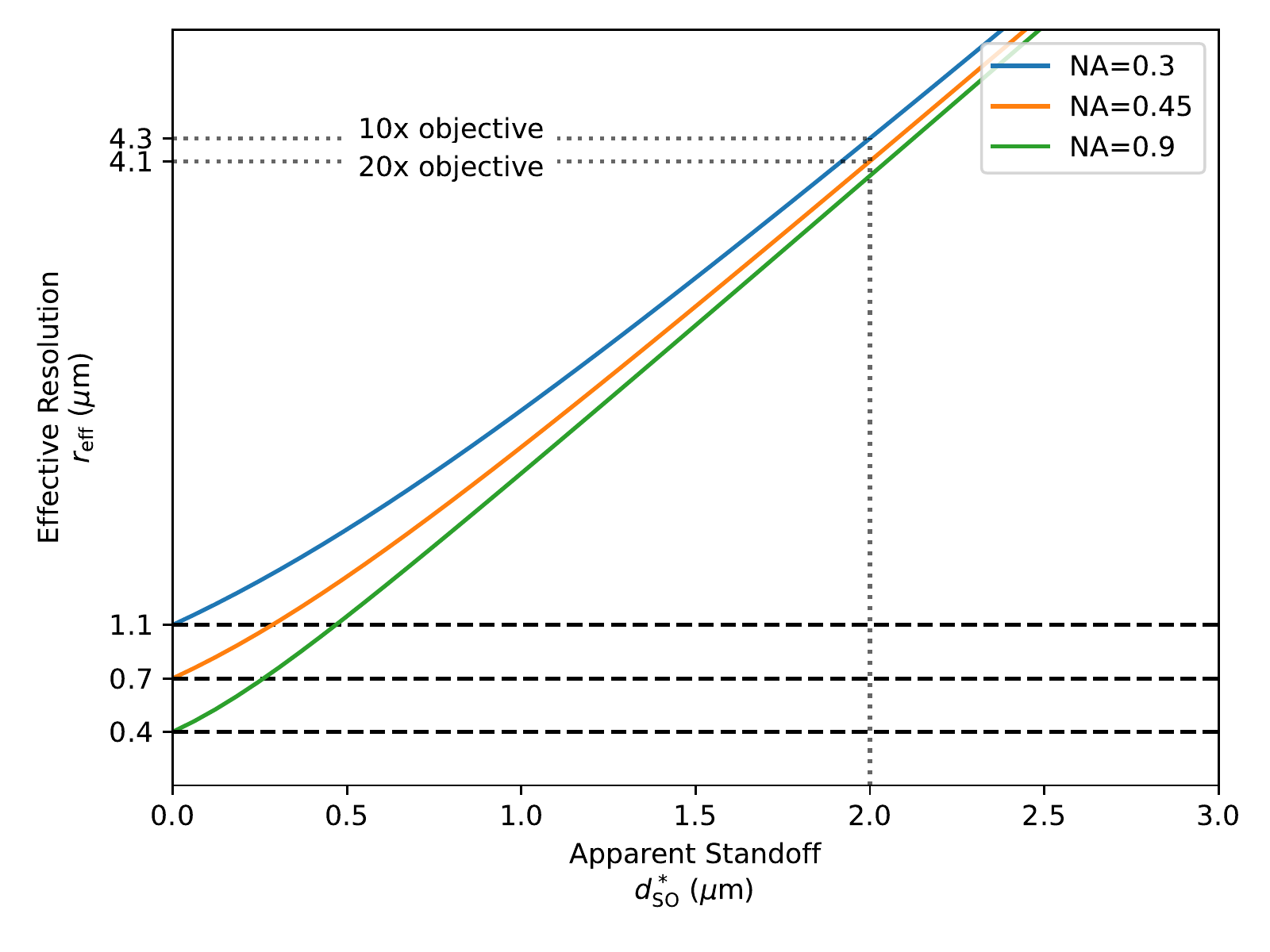}
    \caption{Effective resolution vs apparent standoff (or actual standoff, neglecting NV layer thickness). Black dashed horizontal lines indicate diffraction limited resolution. Grey dotted lines roughly correspond to the situation for the measurements shown in the main text.}
    \label{fig:effectivestandoff}
\end{figure}

\section{Optical Profilometry}

As discussed in the main text, an obvious requirement for standoff minimization is that the sample and diamond surfaces are flat and free of protrusions. We characterized both the diamond and sample using optical profilometry. Note that (without further analysis) profilometry assumes the same material across the image, whereas for the sample there are different materials present, such as the large damaged region where the silicon substrate is exposed. Nevertheless, the protrusions from both surfaces appear to be sufficiently small $\leq 0.5\,\mu{\rm m}$ compared with the standoff distances measured, indicating polishing and cleanliness were not limiting in our case. This detail will require further attention in order to reach the diffraction limit, which requires $\sim100\,{\rm nm}$ standoff. 

\begin{figure}
    \centering
    \includegraphics[width=1.0\textwidth]{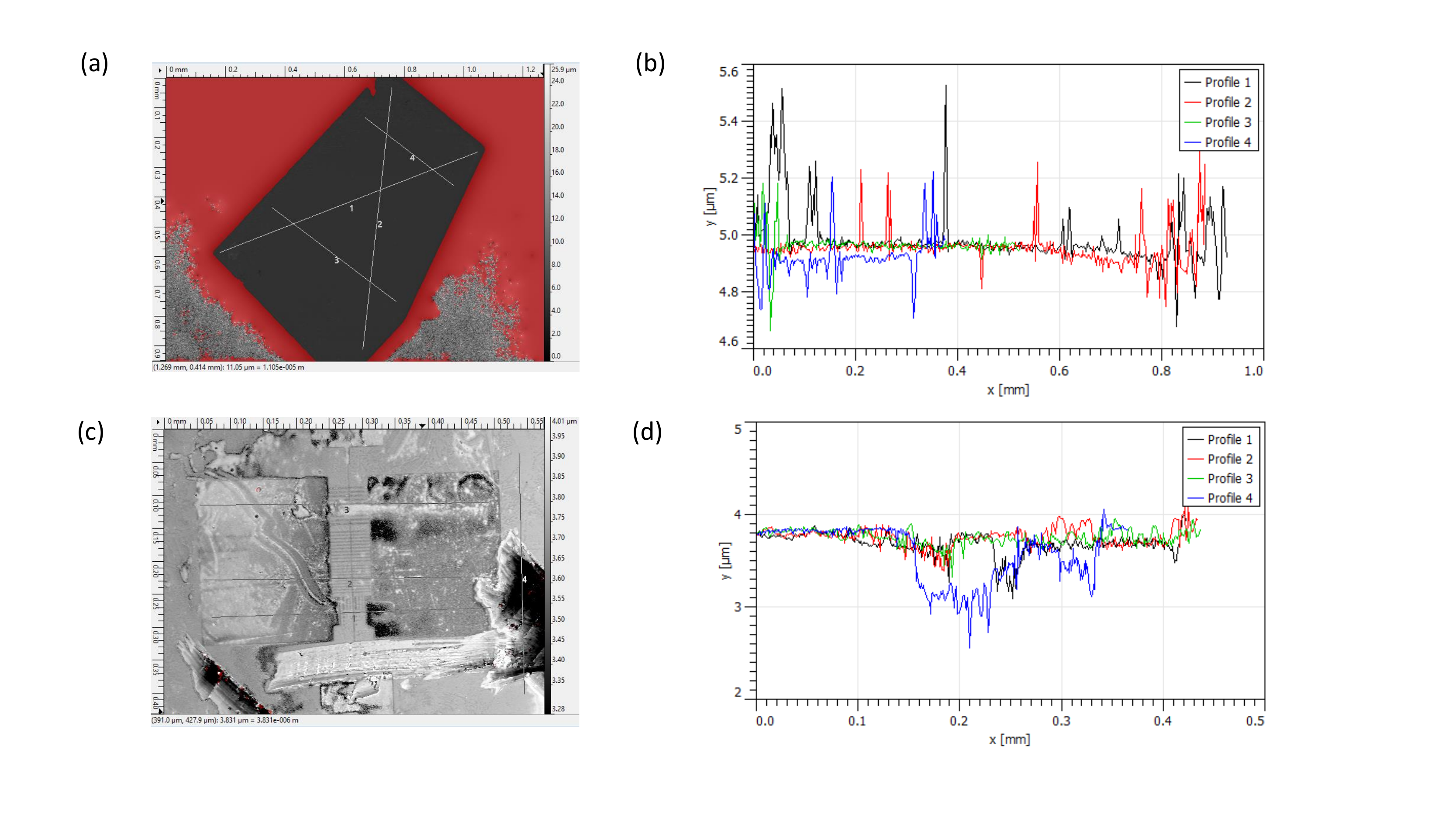}
    \caption{(a) Topographic image of the diamond used in the widefield probe. (b) line-cuts of the diamond surface, indicating only protrusions less than $0.5\,\mu{\rm m}$ (c) Topographic image of the magnetic device. (d) line-cuts of the device surface, indicating only protrusions less than $0.3\,\mu{\rm m}$ above the surface.}
    \label{fig:profiles}
\end{figure}

\end{widetext}

\end{document}